# Penalized versus constrained generalized eigenvalue problems

Irina Gaynanova*, James G. Booth† and Martin T. Wells‡

**Abstract**

We investigate the difference between using an $\ell_1$ penalty versus an $\ell_1$ constraint in generalized eigenvalue problems arising in multivariate analysis. Our main finding is that the $\ell_1$ penalty may fail to provide very sparse solutions; a severe disadvantage for variable selection that can be remedied by using an $\ell_1$ constraint. Our claims are supported both by empirical evidence and theoretical analysis. Finally, we illustrate the advantages of the $\ell_1$ constraint in the context of discriminant analysis and principal component analysis.

## 1 Introduction

Modern technologies allow to collect thousands and millions of measurements, however often only a small subset of these measurements is relevant for the scientific question of interest. As such, variable selection has been a prominent theme in the statistics literature George (2000). One of the common variable selection techniques is $\ell_1$ penalization, which was first used in linear regression. Given the least squares loss function $f(\beta)$, Tibshirani (1996) propose to estimate the parameter vector $\beta$ as

$$\hat{\beta}_\lambda = \arg\min_\beta \{f(\beta) + \lambda\|\beta\|_1\}. \tag{1}$$

The $\ell_1$ penalty in (1) is motivated by the dual optimization problem

$$\hat{\beta}_\tau = \arg\min_\beta \{f(\beta)\} \quad \text{subject to} \quad \|\beta\|_1 \le \tau. \tag{2}$$

Geometrically, the $\ell_1$ constraint in the dual problem (2) projects the solution vector onto the polytope thus forcing some components of $\hat{\beta}_\tau$ to be exactly zero (Figure 1). The constraint tuning parameter $\tau$ controls the size of the polytope, with the number of nonzero variables in $\hat{\beta}_\tau$ increasing with the values of $\tau$. The convexity of least squares function $f(\beta)$ ensures the equivalence between the solutions to (1) and (2) (Bertsekas 1999, Proposition 5.2.1). In particular, for all $\tau \geq 0$ there exists $\lambda \geq 0$ such that $\hat{\beta}_\lambda = \hat{\beta}_\tau$, and vice versa. Since $\hat{\beta}_\lambda = \hat{\beta}_\tau$ implies $f(\hat{\beta}_\lambda) - f(\hat{\beta}_\tau) = 0$, it is said that (1) and (2) have zero duality gap or the strong duality holds (Boyd & Vandenberghe 2004, Chapter 5). The estimator $\hat{\beta}_\lambda$ has nice theoretical properties, see Bickel et al. (2009), Wainwright (2009), Zhao & Yu (2006) and references therein. As a result, the $\ell_1$ penalty has since been used in many other convex problems in statistics Friedman et al. (2008), Mai et al. (2012), Meinshausen & Bühlmann (2006), Rothman (2012).

---

*Corresponding author. Email: ig93@cornell.edu. Mailing address: Cornell University, Department of Statistical Science, 1173 Comstock Hall, Ithaca, NY 14853, USA.

†Department of Biological Statistics and Computational Biology, Cornell University, Ithaca,NY

‡Department of Statistical Science, Cornell University, Ithaca, NY

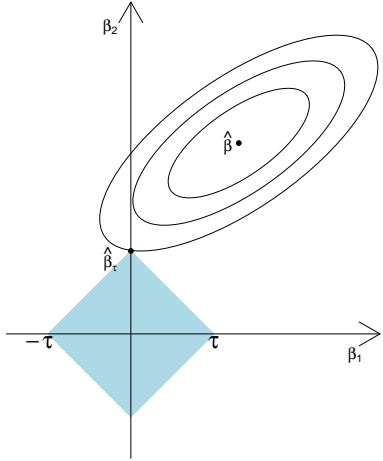

Figure 1: Estimation picture for $\ell_1$ constrained linear regression. Shown are contours of the least squares function $f(\beta)$ and constraint $\|\beta\|_1 \leq \tau$.

Given the success of $\ell_1$ penalty as a variable selection tool in convex problems, it has also been used in nonconvex problems Allen et al. (2014), Allen & Tibshirani (2010), Bien & Tibshirani (2011), Shen & Huang (2008), Witten & Tibshirani (2011), Zou et al. (2006). Some methods consider the dual $\ell_1$ constraint formulation Jolliffe et al. (2003), Witten et al. (2009). The relative popularity of $\ell_1$ penalty over $\ell_1$ constraint is largely due to computational reasons. Usually, it is faster to solve $\ell_1$-penalized problems than $\ell_1$-constrained problems. While the two approaches are equivalent for convex problems, they are not necessarily equivalent for nonconvex problems. In particular, some solutions to $\ell_1$-constrained problem may not be obtained by $\ell_1$-penalized problem, leading to a nonzero duality gap. To our knowledge, this fact has been largely ignored in the statistics literature. Therefore, the effect of this discrepancy on the variable selection performance of $\ell_1$ penalty remains unknown.

The purpose of this work is to understand the difference between the $\ell_1$ penalty and the $\ell_1$ constraint in the subclass of nonconvex problems, generalized eigenvalue problems (GEP). These problems are common in multivariate analysis, and our choice has been largely motivated by the empirical variable selection performance of penalized discriminant analysis Witten & Tibshirani (2011). It has been observed in simulations Mai et al. (2012) and data applications Witten & Tibshirani (2011) that the method consistently selects a much larger number of variables than the competitors, sometimes more than 90% of the original set. No sound explanation has been given for this phenomenon.

In this paper we demonstrate that $\ell_1$-penalized GEPs, such as Witten & Tibshirani (2011), select a large number of variables because it is impossible to select fewer variables. Specifically, the solution to $\ell_1$-penalized GEP is restricted to have at least $M$ nonzero variables independently of the choice of the tuning parameter. The value of $M$ can vary significantly depending on the parameters of optimization problem, however our empirical studies indicate that $M$ generally grows with the total number of variables $p$. Given that the $\ell_1$ penalty is primarily motivated by the need of strong variable selection in large $p$ problems, we conclude that $\ell_1$ penalty fails at this task for generalized eigenvalue problems.

Fortunately, the restriction on the number of nonzero variables does not apply to the $\ell_1$ constraint. By varying the tuning parameter $\tau$ of $\ell_1$-constrained GEP, it is possible to obtain a solution with an arbitrarily small level of sparsity. We show that this discrepancy in variable selection performance of $\ell_1$ penalty and $\ell_1$ constraint is due to the nonzero duality gap between the corresponding optimization problems. As such, $\ell_1$ constraint is superior to $\ell_1$ penalty in terms of variable selection in nonconvex problems.

To our knowledge, this is the first work in model selection that recognizes and quantifies the difference between the $\ell_1$ penalty and the $\ell_1$ constraint in nonconvex settings. In particular, we show

1. the existence of a duality gap between the $\ell_1$-penalized and the $\ell_1$-constrained generalized eigenvalue problem;

2. the existence of a lower bound $M$ on the number of variables selected by the $\ell_1$-penalized generalized eigenvalue problem;

3. the superiority of $\ell_1$ constraint over $\ell_1$ penalty as a variable selection tool in nonconvex problems.

The rest of the paper is organized as follows. Section 2 describes the generalized eigenvalue problem. Section 3 demonstrates empirically that $\ell_1$-penalized GEP cannot obtain very sparse solutions. Section 4 provides a lower bound on the number of non-zero components in the $\ell_1$-penalized solution and Section 5 demonstrates that this bound is due to the duality gap between $\ell_1$-penalized and $\ell_1$-constrained optimization problems. Section 6 illustrates the advantage of $\ell_1$ constraint using discriminant analysis and principal component analysis. Section 7 discusses potential directions for future work.

# 2 Generalized eigenvalue problem

Let $Q \in \mathbb{R}^{p \times p}$ (for quadratic function) and $C \in \mathbb{R}^{p \times p}$ (for constraint) be two symmetric, semi positive-definite matrices. In addition, let $C$ be strictly positive-definite. Consider the optimization problem:

$$v = \arg\max_{v \in \mathbb{R}^p} \left\{ v^\top Q v \right\} \quad \text{subject to} \quad v^\top C v \leq 1. \tag{3}$$

Problem (3) is called the generalized eigenvalue problem Van Loan (1976), since the maximum is achieved when $v$ is the leading eigenvector of matrix $C^{-1}Q$. This optimization problem arises in many multivariate statistical problems, including principal component analysis and discriminant analysis:

1. Let $X \in \mathbb{R}^{n \times p}$ be the centered data matrix. The first principal component loading $v$ is defined as
$$v = \arg\max_{v \in \mathbb{R}^p} \left\{ \frac{1}{n} v^\top X^\top X v \right\} \quad \text{subject to} \quad v^\top v \leq 1.$$
Here $Q = \frac{1}{n} X^\top X$ and $C = I$.

2. Let $(X_i, Y_i)$, $i = 1, .., n$, be independent pairs with $X_i \in \mathbb{R}^p$ and $Y_i \in \{1, .., G\}$, where $G$ is the number of classes. Consider the observed within-group sample covariance matrix $W = \frac{1}{N} \sum_{g=1}^{G} (n_g - 1) S_g$, where $S_g$ is the sample covariance matrix for group $g$, and the between-group sample covariance matrix $B = \frac{1}{N} \sum_{g=1}^{G} n_g (\bar{x}_g - \bar{x})(\bar{x}_g - \bar{x})^\top$ with $\bar{x} = \frac{1}{n} \sum_{i=1}^{n} x_i$ and $\bar{x}_g = \frac{1}{n_g} \sum_{i=1}^{n} x_i \mathbb{I}(y_i = g)$. The first discriminant vector $v$ is defined as:
$$v = \arg\max_{v \in \mathbb{R}^p} \left\{ v^\top B v \right\} \quad \text{subject to} \quad v^\top W v \leq 1.$$
Here $Q = B$ and $C = W$.

Canonical correlation analysis can also be viewed as a generalized eigenvalue problem, for details see Witten et al. (2009).

In recent years there has been a lot of interest in extending traditional multivariate analysis methods to high-dimensional settings by enforcing sparsity in the solution vector $v$. A common approach to achieve this goal is to restrict the $\ell_1$ norm of $v$ by modifying (3), either by penalizing the objective function or by adding the $\ell_1$ norm constraint. We define the $\ell_1$-penalized problem (3) as

$$v_\lambda = \arg\max_{v \in \mathbb{R}^p} \left\{ v^\top Q v - \lambda \|v\|_1 \right\} \text{ subject to } v^\top C v \leq 1. \tag{4}$$

Here $\lambda \geq 0$ is the tuning parameter and the $\ell_1$ norm is part of the objective function. In the following section, we demonstrate empirically that $\ell_1$-penalized problem can fail to select a sparse subset of variables.

**Algorithm 1** Optimization algorithm for $\ell_1$-penalized problem with $C = I$.

---
Given: $\lambda > 0$, $Q$, $\varepsilon > 0$, $k_{\max}$
  $v^{(0)} \leftarrow$ dominant eigenvector of $Q$
  $v^{(0)} \leftarrow v^{(0)}/\sqrt{(v^{(0)})^\top v^{(0)}}$
  $k \leftarrow 1$
  **repeat**
    **for** $l \in \{1, ..., p\}$ **do**
      $v_l^{(k)} \leftarrow \text{sign}\left((Qv^{(k-1)})_l\right) \left(|(Qv^{(k-1)})_l| - \lambda/2\right)_+$
    **end for**
    **if** $\{v^{(k)} \neq 0\}$ **then**
      $v^{(k)} \leftarrow v^{(k)}/\sqrt{(v^{(k)})^\top v^{(k)}}$
    **end if**
    $f(v^{(k)}) \leftarrow (v^{(k)})^\top Q v^{(k)} - \lambda \|v^{(k)}\|_1$
    **if** $\{f(v^{(k)}) < 0\}$ **then**
      $v^{(k)} \leftarrow 0$
      $f(v^{(k)}) \leftarrow 0$
    **end if**
    $k \leftarrow k + 1$
  **until** $k = k_{\max}$ or $v^{(k)}$ satisfies $|f(v^{(k)}) - f(v^{(k-1)})| < \varepsilon$.

---

## 3 Empirical evidence for restriction on solution sparsity

For clarity of exposition, we only consider the case $C = I$. Problem (4) simplifies to

$$v_\lambda = \arg\max_{v \in \mathbb{R}^p} \left\{ v^\top Q v - \lambda \|v\|_1 \right\} \text{ subject to } v^\top v \leq 1. \tag{5}$$

We use Algorithm 1 to find the local solution to problem (5). The full derivation of the algorithm is presented in the Appendix.

First, we consider the following synthetic scenarios:

1. $p \in \{500, 2000\}$, $\text{rank}(Q) = 1$ with eigenvalue $\gamma = 1$ and the dominant eigenvector $l$ with components $l_i$ coming from the uniform distribution on $[0, 1]$, standardized as $l^\top l = 1$.

2. $p \in \{500, 2000\}$, $\text{rank}(Q) = 50$, where $Q$ is the sample covariance matrix of 50 observations $x_i$ with $x_{ij} \sim N(0, 1)$ for $j = 1, ..., p$.

Figure 2 illustrates that the number of selected variables decreases when the value of $\lambda$ increases. What is surprising, however, is the sudden drop to zero which is observed in 3 out of 4 cases and is most severe when $p = 2000$ and $\text{rank}(Q) = 1$. Based on Figure 2, it is impossible to select fewer than 1000 features in this scenario. It appears there exists a $\lambda_0$ such that for all $\lambda < \lambda_0$ the solution has at least $M$ non-zero components, and for all $\lambda \geq \lambda_0$ the solution is exactly zero.

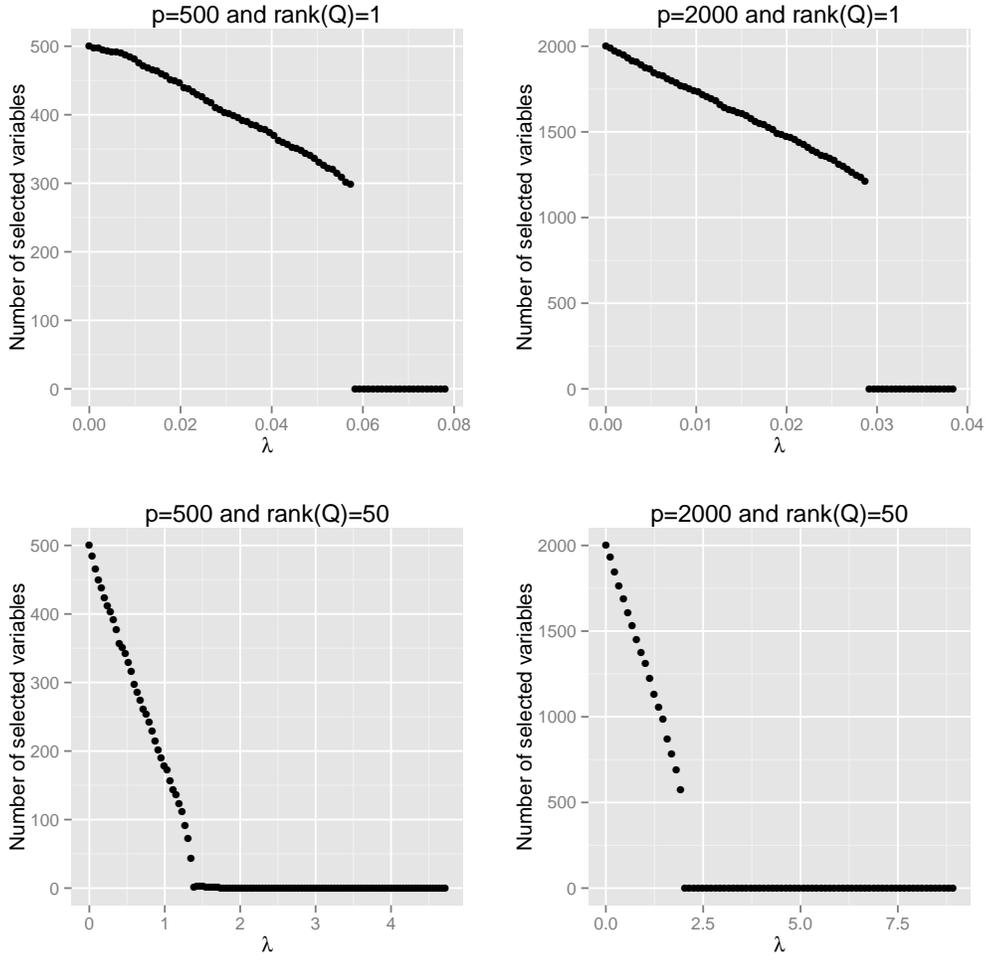

Figure 2: Number of selected variables versus the tuning parameter $\lambda$.

While the convergence to the global solution $v_\lambda$ is not guaranteed due to nonconvexity, the observed empirical behavior of local solution is consistent with the theoretical results of Section 4. We have also investigated the empirical behavior of global solution by applying Algorithm 1 with $r = 100$ random initializations of $v^{(0)}$ for each of the synthetic scenarios. For each initialization, we draw $x_i$, $i = 1, ..., p$, from $N(0, 1)$ and set $v^{(0)} = x/\|x\|_2$. Thus, for each $\lambda$ value we obtain a sequence of $r + 1$ local solutions $v_\lambda, v_{\lambda;1}, ..., v_{\lambda;r}$. Figure 3 illustrates the number of nonzero components in $v_\lambda, v_{\lambda;1}, ..., v_{\lambda;r}$ versus the tuning parameter $\lambda$. The black dots correspond to $v_{\lambda;*}$ with the largest value of the objective function $f(v) = v^\top Q v - \lambda \|v\|_1$, and the grey dots correspond to the remaining $r$ solutions. The sparsity pattern of $v_{\lambda;*}$ is identical to Figure 2. Given the similarity of the results and the additional computational cost associated with random initializations, we restrict the subsequent empirical analysis to the local solution of Algorithm 1.

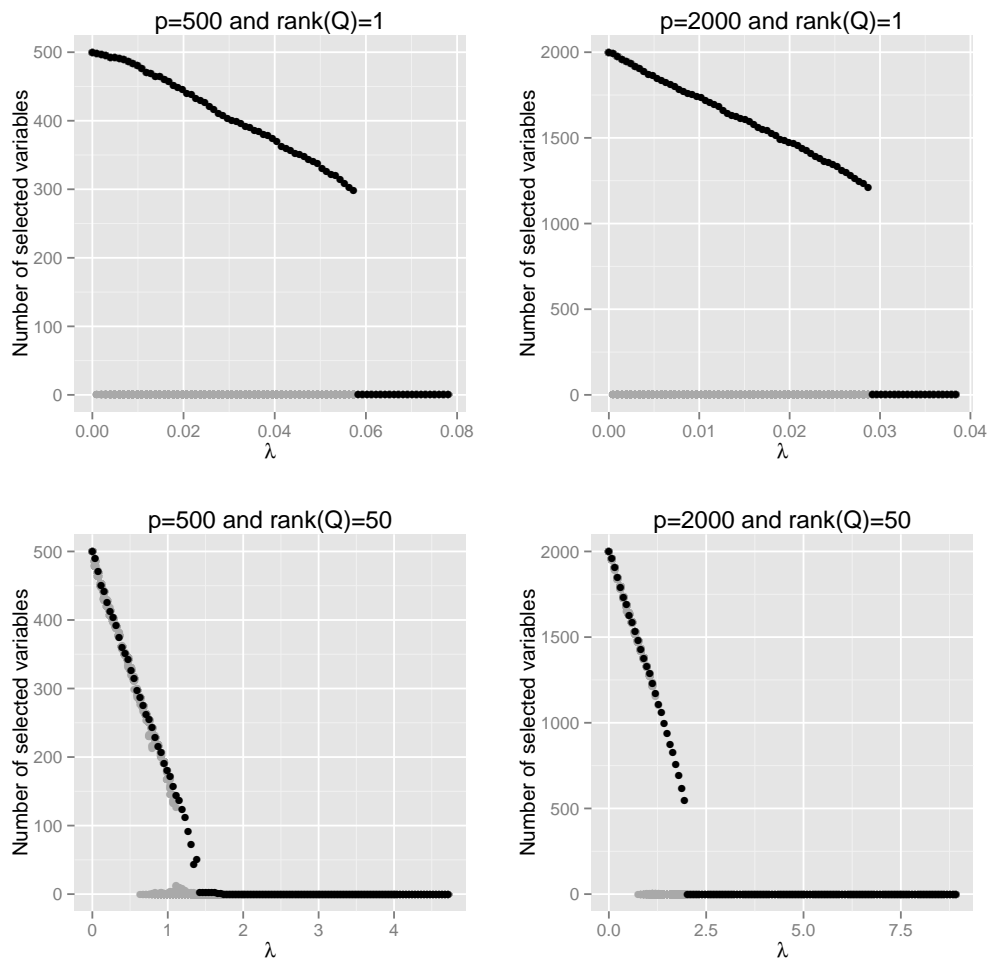

Figure 3: Number of selected variables versus the tuning parameter $\lambda$, black corresponds to the best solution out of 101 initializations of $v^{(0)}$ and grey corresponds to the remaining solutions.

Next, we consider colon cancer dataset Alon et al. (1999) and 14 cancer dataset Ramaswamy et al. (2001). We have chosen these two datasets as they are publicly available and have been extensively studied in the literature Witten & Tibshirani (2011), Zou et al. (2006). Colon cancer dataset is available from http://genomics-pubs.princeton.edu/oncology/affydata/index.html. It contains the expression of 2000 genes from 40 tumor tissues and 22 normal tissues. 14 cancer dataset is available from http://statweb.stanford.edu/~tibs/ElemStatLearn/. It contains 16063 gene expression measurements collected on 198 samples from 14 cancer classes. For the analysis, we select 144 samples that are designated as the training set. Following the recommendation of (Hastie et al. 2009, p. 654), we standardize the data to have mean zero and standard deviation one for each patient. To reduce computational costs, we only consider 2684 genes that have standard deviation above 0.45.

We apply penalized linear discriminant analysis (LDA) and penalized principal components analysis (PCA) to both datasets, more details are provided in Section 6.1. Thus, there are four scenarios:

1. Penalized LDA on colon cancer dataset, $p = 2000$, $\text{rank}(Q) = 1$.

2. Penalized PCA on colon cancer dataset, $p = 2000$, $\text{rank}(Q) = 61$.

3. Penalized LDA on 14 cancer dataset, $p = 2684$, $\text{rank}(Q) = 13$.

4. Penalized PCA on 14 cancer dataset, $p = 2684$, $\text{rank}(Q) = 141$.

Figure 4 shows the number of selected variables versus the tuning parameter for colon cancer and 14 cancer datasets. As in the case with synthetic data, there is a sudden drop to zero as the tuning parameter increases. This behavior appears to be especially problematic with penalized PCA, making it impossible to select less than 1000 variables for either dataset.

These results demonstrate that the $\ell_1$-penalized method is not effective as a variable selection tool. The large number of selected variables makes it impossible to interpret the results or further validate the variables in a lab setting. In Section 4 we demonstrate that this behavior is not an artifact of the chosen optimization algorithm, but rather is intrinsic to penalized generalized eigenvalue problem (4).

## 4  Lower bound on the number of non-zero components

We start by deriving an upper bound on the objective function in (4) for $v$ such that $\|v\|_0 \leq k$.

**Proposition 1.** *Let $q_i$ be the ith row of $Q$ and let $q_i(j)$ be the subvector of $q_i$ of length $j$ with the maximal $\ell_2$ norm. Then*

$$\max_{\|v\|_0 \leq k, v^\top C v \leq 1} \left\{ v^\top Q v - \lambda \|v\|_1 \right\} \leq \frac{\|\tilde{q}(k)\|_2}{\sigma_{\min}(C)},$$

*where $\tilde{q}_i = \max\left(\|q_i(k)\|_2 - \lambda \sqrt{\sigma_{\min}(C)}, 0\right)$.*

The upper bound grows with the value of $k$. In particular, if $\|\tilde{q}(l)\|_2 = 0$, then $\|\tilde{q}(m)\|_2 = 0$ for all $m < l$. As such, we can derive the lower bound on the number of nonzero components in $v_\lambda$.

**Corollary 1.** *Let $q_i$ be the ith row of $Q$ and let $q_i(j)$ be the subvector of $q_i$ of length $j$ with the maximal $\ell_2$ norm. Let $m_\lambda = j_{\min} \in \{1, ..., p\}$ such that $\max_i \|q_i(j)\|_2 > \lambda \sqrt{\sigma_{\min}(C)}$. Then*

$$v_\lambda = 0 \quad or \quad \|v_\lambda\|_0 \geq m_\lambda.$$

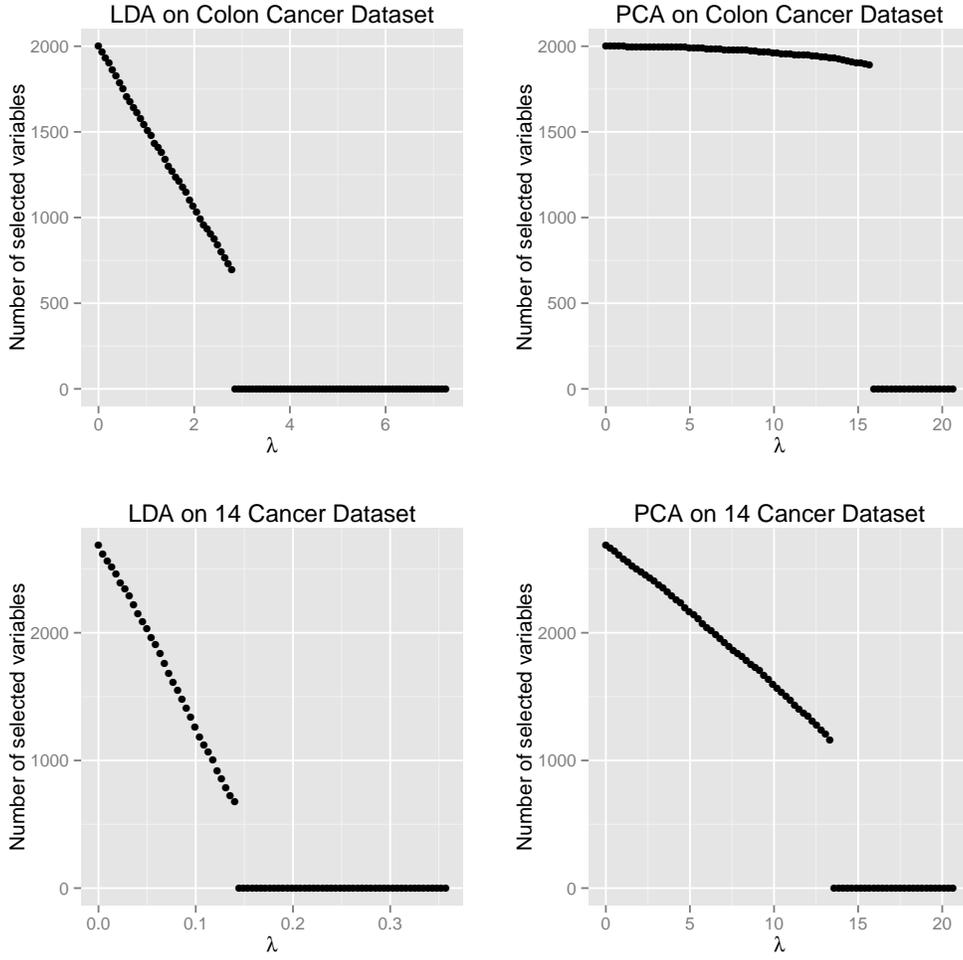

Figure 4: Number of selected variables versus the tuning parameter $\lambda$.

We can further use this result to derive a value of $\lambda_{\max}$.

**Corollary 2.** *Let $q_i$ be the ith row of $Q$ and let $\lambda_{\max} = \max_i \|C^{-1/2} q_i\|_2$. Then for all $\lambda \geq \lambda_{\max}$, $v_\lambda = 0$.*

Since the bound of Proposition 1 applies to any $v$ in the feasible region with $\|v\|_0 \leq k$, the results of Proposition 1 and Corollaries 1 and 2 also apply to the solution of Algorithm 1.

Figure 5 demonstrates the bound from Corollary 1 as a function of $\lambda$ using the examples of Section 3. As the tuning parameter $\lambda$ increases, so does the value of $m_\lambda$ (dashed line on Figure 5). On the other hand, as the tuning parameter $\lambda$ increases, so does the weight of the penalty in the objective function of (4) leading to the smaller number of selected variables (dots on Figure 5). A perfect prediction of the minimal number of nonzero variables requires the dashed line to take the same value as dotted line at $\lambda = \lambda_0$ where the drop happens. For example, when $p = 500$ and $\text{rank}(Q) = 1$, $\lambda_0 \approx 0.06$, $m_{\lambda_0} \approx 150$, whereas the drop happens at $\approx 300$ variables. This discrepancy between the actual minimal number of selected variables and the value of $m_{\lambda_0}$ is not surprising, since the value of $m_\lambda$ is based on an upper bound of the objective function in (4). While this bound appears to be somewhat crude for the case when $\text{rank}(Q) = 50$, it predicts at least 150 features for $p = 500$, $\text{rank}(Q) = 1$ case and at least 500 features for $p = 2000$, $\text{rank}(Q) = 1$ case. Similarly, it predicts at least 34 selected variables for penalized LDA on 14 cancer dataset, at least 267 selected variables for penalized PCA on 14 cancer dataset, at least 38 selected variables for penalized LDA on colon cancer dataset and at least 728 selected variables for penalized PCA on colon cancer dataset.

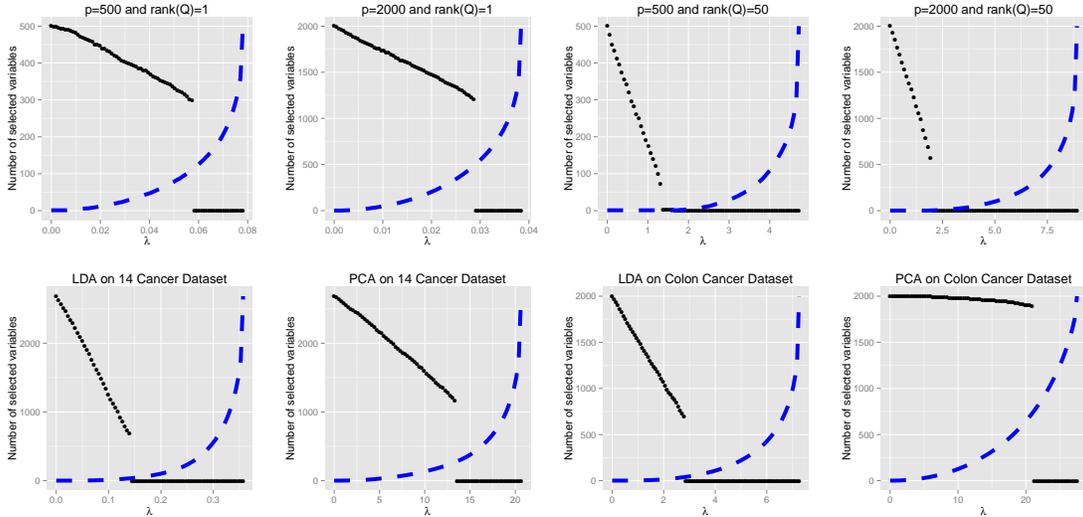

Figure 5: Number of non-zero variables obtained empirically versus the tuning parameter $\lambda$, the dashed line shows the value of $m_\lambda$ from Corollary 1.

## 5  $\ell_1$ penalty versus $\ell_1$ constraint

In Sections 3 and 4 we demonstrated both empirically and theoretically that $\ell_1$-penalized generalized eigenvalue problem can fail to obtain very sparse solutions. Consider $\ell_1$-constrained generalized eigenvalue problem

$$v_\tau = \arg\max_{v \in \mathbb{R}^p} \left\{ v^\top Q v \right\} \quad \text{subject to} \quad v^\top C v \leq 1, \|v\|_1 \leq \tau. \tag{6}$$

Here $\tau \geq 0$ is a tuning parameter which constrains the $\ell_1$ norm of $v$. It is natural to ask whether the solutions to problems (4) and (6) are the same, and whether the same restriction on solution sparsity applies to (6). A partial answer to this question is given in Proposition 2.

**Proposition 2.** *For every $\lambda \geq 0$ there exists $\tau \geq 0$ such that $v_\lambda = v_\tau$.*

The reverse is true for convex problems such as LASSO (Bertsekas 1999, Proposition 5.2.1), however the generalized eigenvalue problem is nonconvex. Following (Bertsekas 1999, Chapter 5) and (Boyd & Vandenberghe 2004, Chapter 5.3), we use a geometry-based approach to visualize the relationship between the solutions to the $\ell_1$-constrained and the $\ell_1$-penalized optimization problems in the following example.

*Example:* Let $p = 2$, $C = I$ and $\text{rank}(Q) = 1$. Let $\gamma = 1$ be the positive eigenvalue of $Q$ and $l$ be the corresponding eigenvector, so that $Q = \gamma ll^\top$. We consider two scenarios:

1. $x = (0.2, 0.8)^\top$, $l = x/\|x\|_2$;

2. $x = (0.5, 0.6)^\top$, $l = x/\|x\|_2$.

The corresponding $\ell_1$-constrained optimization problem (6) becomes

$$v_\tau = \arg\min_{v \in \mathbb{R}^p} -(v^\top l)^2 \text{ subject to } v^\top v \leq 1, \|v\|_1 \leq \tau. \tag{7}$$

This problem defines the set $S$ of constrained pairs

$$S = \left\{(h, f) \,\middle|\, h = \|v\|_1, f = -(v^\top l)^2 \text{ for all } v \in \mathbb{R}^p, v^\top v \leq 1\right\}. \tag{8}$$

Using the set $S$, (7) can be viewed as a minimal common point problem: finding a point $(h', f')$ with a minimal $f$-coordinate among the points common to set $S$ and halfspace $h \leq \tau$,

$$\left\{(h', f') \in S \,\middle|\, f' = \min_{(h,f) \in S, h \leq \tau} f\right\}. \tag{9}$$

By definition of $v_\tau$, $f' = -(v_\tau^\top l)^2$ and $h' = \|v_\tau\|_1$. We construct the corresponding sets $S$ for both scenarios in Figure 6 and identify the minimal common point using $\tau = 1.1$.

Consider the corresponding $\ell_1$-penalized optimization problem (4):

$$v_\lambda = \arg\min_{v \in \mathbb{R}^p} -(v^\top l)^2 + \lambda \|v\|_1 \text{ subject to } v^\top v \leq 1. \tag{10}$$

Using the set $S$ in (8), we can view (10) as finding the point $(h'', f'') \in S$ such that

$$(h'', f'') = \arg\min_{(h,f) \in S} \{f + \lambda h\}.$$

By definition of $v_\lambda$, $h'' = \|v_\lambda\|_1$ and $f'' = -(v_\lambda^\top l)^2$.

The solutions to (7) and (10) are the same if $(h', f') = (h'', f'')$. This occurs when $f = -\lambda h$ is the supporting hyperplane to the set $S$ at the point $(h', f')$. Figure 6 shows whether such a hyperplane can be constructed in both scenarios. In the first scenario the hyperplane can be constructed for each $\tau \geq 1$, and in particular for $\tau = 1.1$. In the second scenario, the hyperplane cannot be constructed for $\tau = 1.1$ as it has to lie below the point $(0,0)$ and the minimal point of $S$ corresponding to $h = 1.4$. Moreover, this is true not only for $\tau = 1.1$ but for all values of $\tau$ between 1 and 1.4. Hence, for these $\tau$ there exists no $\lambda$ such that $v_\lambda = v_\tau$.

Consider the shape of the set $S$ in the second scenario. For all $\tau < 1.4$, $(h', f') = (0, 0)$ is the only point at which it is possible to construct the supporting hyperplane to the set $S$. This implies $h'' = \|v_\lambda\|_1 = 0$, hence $v_\lambda = 0$ is the corresponding solution to the dual problem

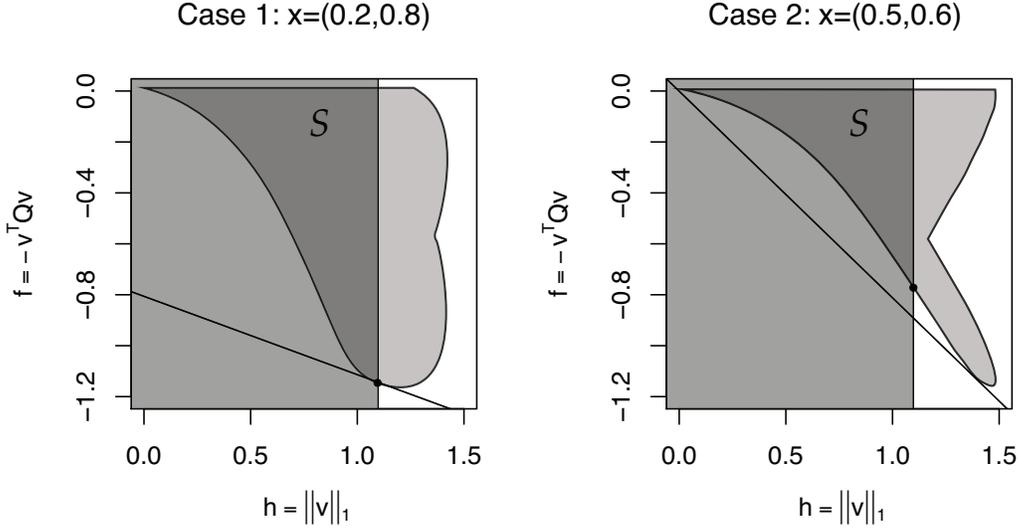

Figure 6: Visualization of the set $S$, the minimum common point of $S$ and $h \leq 1.1$, and the supporting hyperplane for the set $S$. The eigenvector of matrix $Q$ is equal to $l = x/\|x\|_2$.

(10) for all $\tau < 1.4$. In contrast, $v_\tau = 0$ only for $\tau = 0$. Therefore there exists no $\lambda \geq 0$ such that $\|v_\lambda\|_1 = \tau$ for $\tau \in (0, 1.4)$, leading to a constraint on the sparsity level of the solution $v_\lambda$.

In the language of optimization theory, the Lagrangian dual problem defines the supporting hyperplane to $S$ in (8), and hence the optimal (primal) solution is greater than the dual solution (weak duality). If the supporting hyperplane intersects $S$ at a single point, as in scenario one above, the optimization problem is said to have the zero duality gap (strong duality) property. If the objective function is convex, as in the LASSO Tibshirani (1996), strong duality is guaranteed by Slaters constraint (Boyd & Vandenberghe 2004, Chapter 5). Unlike the LASSO, (3) is not a convex problem and therefore this guarantee no longer applies.

Our example demonstrates the existence of a duality gap between problems (4) and (6); there exist values of $\tau > 0$ such that the solution $v_\tau$ cannot be obtained by solving (4). Moreover, these unattainable values of $\tau$ correspond to sparse solutions, $v_\tau$ with very few non-zero components. Therefore, there is a restriction on the sparsity of solutions obtained by solving the $\ell_1$-penalized problem (4), but there is no restriction on the sparsity of the solutions obtained by solving the corresponding $\ell_1$-constrained problem (6).

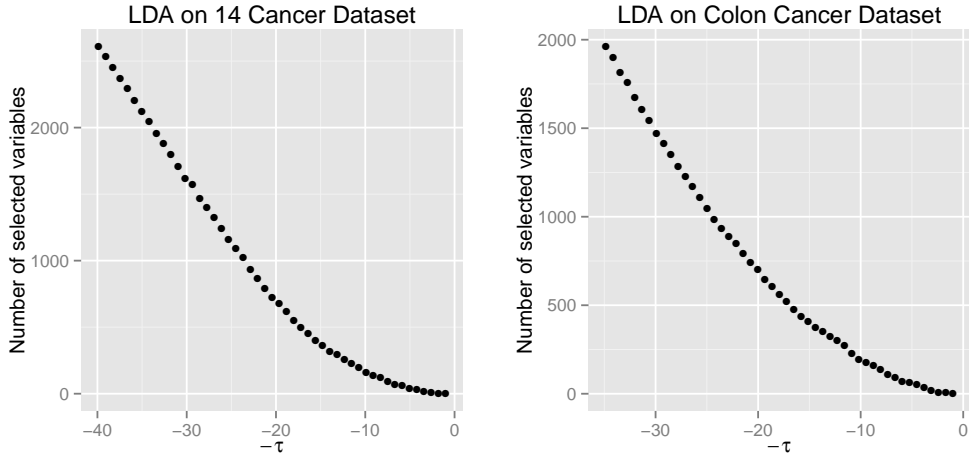

Figure 7: Number of selected variables versus the tuning parameter for the $\ell_1$-constrained LDA.

# 6 Variable selection with $\ell_1$ constraint

## 6.1 Fisher's Linear Discriminant Analysis

Let $(X_i, Y_i)$, $i = 1, \ldots, n$, be independent pairs with $X_i \in \mathbb{R}^p$ and $Y_i \in \{1, \ldots, G\}$, where $G$ is the number of classes. Let $W$ and $B$ be the within-group sample covariance matrix and the between-group sample covariance matrix respectively. Further, assume that $X$ is scaled so that $\text{diag}(W) = I$. Witten & Tibshirani (2011) find the first penalized discriminant vector as

$$v_\lambda = \arg\max_{v \in \mathbb{R}^p} \left\{ v^\top B v - \lambda \sum_{j=1}^p |v_j| \right\} \text{ subject to } v^\top v \leq 1. \qquad (11)$$

We use (11) for the analysis of the colon cancer dataset Alon et al. (1999) and 14 cancer dataset Ramaswamy et al. (2001). Figure 4 shows the number of non-zero features selected by Algorithm 1 versus the tuning parameter $\lambda$. Empirically it is impossible to select less than 600 variables for colon cancer dataset and less than 500 variables for 14 cancer dataset.

Now consider the constrained version of (11):

$$v_\tau = \arg\max_{v \in \mathbb{R}^p} \left\{ v^\top B v \right\} \text{ subject to } v^\top v \leq 1, \quad \|v\|_1 \leq \tau. \qquad (12)$$

The local solution to Problem (12) can be found using Algorithm 1 with the following modification: for each iteration $k$ choose $\lambda^{(k)}$ such that $\|v^{(k)}_{\lambda^{(k)}}\|_1 = \tau$. Usually, such a $\lambda^{(k)}$ is found by performing a binary search on the grid $[0, \lambda_{\max}]$. Figure 7 shows the number of non-zero components in $v_\tau$ versus the tuning parameter $\tau$. As $\tau$ increases, so does the number of selected variables. Moreover, it is possible to select an arbitrarily small number of variables.

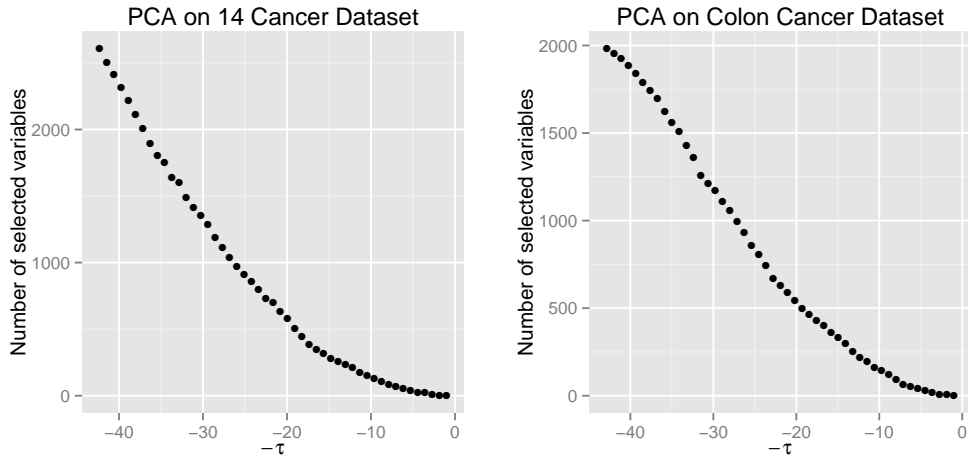

Figure 8: Number of selected variables versus the tuning parameter for the $\ell_1$-constrained PCA.

## 6.2 Principal Component Analysis

Let $X_i \in \mathbb{R}^p$, $i = 1, ..., n$, be independent samples and let $S$ be the sample covariance matrix. Assume that $X$ is scaled so that $S = \frac{1}{n}X^\top X$. The first penalized principal component loading is defined as

$$v_\lambda = \arg\max_{v \in \mathbb{R}^p} \left\{ \frac{1}{n} v^\top X^\top X v - \lambda \sum_{j=1}^p |v_j| \right\} \quad \text{subject to} \quad v^\top v \leq 1. \qquad (13)$$

We use (13) for the analysis of the colon cancer dataset Alon et al. (1999) and 14 cancer dataset Ramaswamy et al. (2001). Figure 4 shows the number of non-zero features selected by Algorithm 1 versus the tuning parameter $\lambda$. Empirically it is impossible to select less than 1700 variables for colon cancer dataset and less than 1000 variables for 14 cancer dataset.

Now consider the constrained version of (13):

$$v_\tau = \arg\max_{v \in \mathbb{R}^p} \left\{ \frac{1}{n} v^\top X^\top X v \right\} \quad \text{subject to} \quad v^\top v \leq 1, \quad \|v\|_1 \leq \tau. \qquad (14)$$

Problem (14) has been previously used for PCA by Jolliffe et al. (2003) and Witten et al. (2009). The local solution to (14) can be found using Algorithm 1 in the same way as the local solution to (12). Figure 8 shows the number of non-zero components in $v_\tau$ versus the tuning parameter $\tau$. With the constrained PCA, it is possible to select an arbitrarily small number of variables.

## 7 Discussion

In this paper we restricted our analysis to a subclass of nonconvex problems, generalized eigenvalue problems. Several methods in the literature use nonconvex functions with $\ell_1$ penalty that are not generalized eigenvalue problems Allen et al. (2014), Allen & Tibshirani (2010), Bien & Tibshirani (2011), Shen & Huang (2008), Zou et al. (2006). We conjecture that the restriction on the solution sparsity is likely present in these methods as well due to the expected nonzero duality gap between the $\ell_1$ penalty and the $\ell_1$ constraint. The restriction on the solution sparsity directly affects the variable selection properties of the corresponding estimators, making them unfavorable in large $p$ scenarios where strong variable selection is desired. To assess the presence of such restriction for a given nonconvex problem, we recommend plotting the number of nonzero variables versus the tuning parameter $\lambda$ as has been done in Figures 2 and 4. In case the drop to zero is observed, we recommend using the $\ell_1$-constrained formulation. In future work, we plan to perform the empirical investigation of the presence of a duality gap for a wider range of nonconvex problems, and generalize the theoretical results of Section 4 to a broader class of nonconvex functions.

# Acknowledgement

We thank Michael Todd for a useful discussion of duality theory. We also thank Daniela Witten for the helpful comments on the earlier drafts of this manuscript. This research was partially supported by NSF grants DMS-1208488 and DMS-0808864

# Appendix

### Derivation of Algorithm 1.

Following Witten & Tibshirani (2011), (5) can be recast as a biconvex optimization problem

$$\text{maximize}_{u,v} \left\{ 2u^\top Q^{1/2} v - \lambda \sum_{j=1}^{p} |v_j| - u^\top u \right\} \text{ subject to } v^\top v \leq 1, \tag{15}$$

since maximizing with respect to $u$ gives $u = Q^{1/2}v$. The problem (15) is convex with respect to $u$ when $v$ is fixed and is convex with respect to $v$ when $u$ is fixed. This property allows the use of Alternate Convex Search (ACS) to find the solution (Gorski et al. 2007, Section 4.2.1). ACS ensures that all accumulation points are partial optima and have the same function value (Gorski et al. 2007, Theorem 4.9).

Starting with an initial value $v^{(0)}$ the algorithm proceeds by iterating the following two steps:

**Step 1** $u^{(k)} = \arg\max_u \left\{ 2u^\top Q^{1/2} v^{(k)} - u^\top u \right\} = Q^{1/2} v^{(k)}$

**Step 2** $v^{(k+1)} = \arg\max_v \left\{ 2(u^{(k)})^\top Q^{1/2} v - \lambda \sum_{j=1}^{p} |v_j| \right\}$ subject to $v^\top v \leq 1$.

Following (Witten & Tibshirani 2011, Proposition 2), it is useful to reformulate Step 2 as

$$q^{(k+1)} = \arg\max_q \left\{ 2(u^{(k)})^\top Q^{1/2} q - \lambda \sum_{j=1}^{p} |q_j| - q^\top q \right\} \tag{16}$$

where, if $q^{(k+1)} = 0$, then $v^{(k+1)} = 0$, else $v^{(k+1)} = q^{(k+1)}/\sqrt{(q^{(k+1)})^\top q^{(k+1)}}$. Since problem (16) is convex with respect to $q$, the solution $q^{(k+1)}$ satisfies KKT conditions Boyd & Vandenberghe (2004)

$$2Q^{1/2}u^{(k)} - 2q^{(k+1)} - \lambda \Gamma = 0, \tag{17}$$

where $\Gamma$ is a $p$-vector and each $\Gamma_j$ is a subgradient of $|q_j^{(k+1)}|$, i.e. $\Gamma_j = 1$ if $q_j^{(k+1)} > 0$, $\Gamma_j = -1$ if $q_j^{(k+1)} < 0$ and $\Gamma_j$ is between $-1$ and $1$ if $q_j^{(k+1)} = 0$. From (17)

$$q_j^{(k+1)} = \text{sign}((Q^{1/2}u^{(k)})_j) \left( |(Q^{1/2}u^{(k)})_j| - \frac{\lambda}{2} \right)_+. \tag{18}$$

Algorithm 1 results from combining Steps 1 and 2 with the update (18).

**Proofs.**

*Proof of Proposition 1.* When $\|v\|_0 \leq k$ and $v^\top C v \leq 1$,

$$v^\top Q v - \lambda \|v\|_1 = \sum_{i=1}^p |v_i| \left( s_i q_i^\top v - \lambda \right) \leq \sum_{i=1}^p |v_i| (\|q_i(k)\|_2 \|v\|_2 - \lambda)$$

$$\leq \sum_{i=1}^p |v_i| \left( \frac{\|q_i(k)\|_2}{\sqrt{\sigma_{\min}(C)}} - \lambda \right) \leq \frac{1}{\sqrt{\sigma_{\min}(C)}} \sum_{i=1}^p |v_i| \tilde{q}_i$$

$$\leq \frac{1}{\sqrt{\sigma_{\min}(C)}} \|\tilde{q}_{(k)}\|_2 \|v\|_2 \leq \frac{1}{\sigma_{\min}(C)} \|\tilde{q}_{(k)}\|_2.$$

□

*Proof of Proposition 2.* Fix any $\lambda \geq 0$ and let $v_\lambda$ be the solution to (4). It follows that for any $v$ such that $v^\top v \leq 1$,

$$v_\lambda^\top Q v_\lambda - \lambda \|v_\lambda\|_1 \geq v^\top Q v - \lambda \|v\|_1. \tag{19}$$

Consider (6) with $t = \|v_\lambda\|_1$. From (19) for each $v$ such that $v^\top v \leq 1$ and $\|v\|_1 \leq t$

$$v_\lambda^\top Q v_\lambda \geq v^\top Q v + \lambda (\|v_\lambda\|_1 - \|v\|_1) = v^\top Q v + \lambda (t - \|v\|_1) \geq v^\top Q v.$$

This means $v_\lambda$ is the solution to (6), hence $v_t = v_\lambda$. □